# Revisiting double diffusion encoding MRS in the mouse brain at 11.7T: which microstructural features are we sensitive to?

Mélissa Vincent [1,2], Marco Palombo [3], Julien Valette [1,2,*]

[1] Commissariat à l'Energie Atomique et aux Energies Alternatives (CEA), MIRCen, F-92260, Fontenay-aux-Roses, France

[2] Neurodegenerative Diseases Laboratory, UMR9199, CEA, CNRS, Université Paris Sud, Université Paris-Saclay, F-92260, Fontenay-aux-Roses, France

[3] Department of Computer Science and Centre for Medical Image Computing, University College of London, London, WC1E 6BT, United Kingdom

**\*Corresponding author**: Dr. Julien Valette, Commissariat à l'Energie Atomique et aux Energies Alternatives (CEA), Direction de la Recherche Fondamentale (DRF), Institut de Biologie François Jacob, MIRCen, 18 Route du Panorama, 92260 Fontenay-aux-Roses, France; Phone: +33 1 46 54 81 30; e-mail: julien.valette@cea.fr

**Abstract**

Brain metabolites, such as N-acetylaspartate or myo-inositol, are constantly probing their local cellular environment under the effect of diffusion. Diffusion-weighted NMR spectroscopy therefore presents unparalleled potential to yield cell-type specific microstructural information. Double diffusion encoding (DDE) relies on two diffusion blocks which relative directions describe a varying angle during the course of the experiment. Unlike single diffusion encoding, DDE measurements at long mixing time display some angular modulation of the signal amplitude which reflects compartment shape anisotropy, while requiring relatively low gradient strength. This angular dependence has been formerly used to quantify cell fiber diameter using a model of isotropically oriented infinite cylinders. However, it has been little explored how additional features of the cell microstructure, such as cell body diameter, fiber length and branching may also influence the DDE signal. Here, we used a cryoprobe as well as state-of-the-art post-processing to perform DDE acquisitions with high accuracy and precision in the mouse brain at 11.7 T. We then compared our results to simulated DDE datasets obtained in various 3D cell models in order to pinpoint which features of cell morphology may influence the most the angular dependence of the DDE signal. While the infinite cylinder model poorly fits our experimental data, we show that incorporating branched fiber structure in our model is paramount to sensibly interpret the DDE signal. Lastly, experiments and simulations in the short mixing time regime suggest that some sensitivity to cell body diameter might be retrieved, although additional experiments would be required to further support this statement.

1. Introduction

Elucidating brain cells' structural complexity is a formidable task that has been driving decades of intense methodological development in the field of nuclear magnetic resonance (NMR) *in vivo*. Magnetic resonance imaging (MRI) is based on the detection of the abundant pool of water molecules residing in biological tissues and offers a variety of contrast sources such as diffusion MRI, which has led to invaluable insights into brain microstructure (Alexander et al., 2019; Novikov et al., 2019). Improvement in scanner hardware has allowed the implementation of magnetic resonance spectroscopy (MRS) at very high field *in vivo*, yielding highly-resolved spectra for the accurate quantification of brain metabolites (Tkac et al., 2009). MRS therefore opens up a direct and specific access to brain biochemistry to non-invasively assess and monitor pathologies affecting the central nervous system (CNS) (Gill et al., 1989). It is furthermore possible to be sensitive to metabolite diffusion, e.g. using a pair of diffusion gradients as proposed by Stejskal-Tanner (Stejskal and Tanner, 1965). This is of particular interest because intracellular metabolites are constantly exploring their local environment under the effect of diffusion, thus potentially reporting precious information on several attributes of the cell microstructure such as cell fiber length or diameter (Palombo et al., 2018c). Moreover, unlike water, some metabolites are known to be confined in specific cell compartments, as in the case of NAA and glutamate (Glu), which are neuronal markers, or myo-inositol (Ins) and choline compounds (tCho), which are associated to glial cells (Choi et al., 2007). This enables probing of the microstructure of specific cell types, which might be highly relevant in a neuropathological context where severe morphological alterations of specific cells can occur, for instance neuronal atrophy or neuroinflammation (Ercan et al., 2016).

Different diffusion-weighted MRS (DW-MRS) approaches have been proposed to quantify cell microstructure, such as high-b value experiments (mainly to probe fiber diameter but also presumably yielding some sensitivity to cell body diameter (Ligneul et al., 2019; Palombo et al., 2017; Palombo M, 2019; Palombo et al., 2018a, b)), or measurements of the apparent diffusion coefficient (ADC) up to very long diffusion times (td) to probe long-range cell structure (potentially enabling quantification of cell fiber branching and length(Ligneul et al., 2019; Palombo et al., 2016)). Such approaches were recently used to quantitatively estimate alterations of astrocytic morphology in a mouse model of reactive astrocytes, where the diffusion of Ins was significantly different compared to the control group, while the diffusion of other metabolites remained unchanged (Ligneul et al., 2019). In that work, modeling of Ins data resulted in higher values for fiber diameter and length in the case of activated astrocytes, which was confirmed by quantitative morphological analysis of astrocytes using confocal microscopy *ex vivo*, thus demonstrating the potential of the technique for microstructure quantification. Unfortunately, going to very long diffusion times or high b-values leads to low signal-to-noise ratio, and high b-value requires very high gradient strength, making the implementation of such measurements on clinical scanners highly challenging.

One approach circumventing these issues is double diffusion encoding (DDE) (Shemesh et al., 2010). In DDE, two diffusion blocks are successively applied while the relative directions of the corresponding diffusion-sensitizing gradients are varied by an angle φ during the course of the experiment. Unlike the more conventional Single Diffusion Encoding (SDE), DDE has been shown to display sensitivity to compartment shape anisotropy (CSA) together with microscopic anisotropy (μA) that arises from the presence of restricting boundaries (Mitra, 1995; Ozarslan, 2009; Ozarslan and Basser, 2008). Theoretical analysis of DDE was proposed by Mitra who examined two regimes based on null/short vs. long mixing time (TM) between the two diffusion blocks (Mitra, 1995). The long TM case is particularly interesting to disentangle between compartments without CSA (e.g. spheres, or a tortuous medium with typical pore size much below the diffusion distance) and with CSA (e.g. ellipsoidal pores or fibers). Indeed, Mitra demonstrated that there is an angular dependence of the NMR signal magnitude for ellipsoidal pores but not for spherical ones. In this regime, in the case of ellipsoidal pores, a signal maximum is consequently expected in the parallel/antiparallel cases (φ=0 and 180°) whereas a signal minimum is predicted for the perpendicular case (φ=90 and 270°).

DDE-MRS was successfully pioneered by Shemesh *et al.* to non-invasively follow microstructural alterations of ischemic tissues in a stroke model (Shemesh et al., 2014). Signal modulation as a function of φ unambiguously demonstrated that metabolite diffusion compartments exhibited CSA. Moreover, the authors reported a dramatic increase of the amplitude of the NMR signal angular dependence 24 hour after the onset of ischemia for NAA, tCho and total creatine (tCr) suggesting that CSA is increasing. Shemesh *et al.* further investigated the potential of DDE-MRS to quantitatively extract cell fiber diameter $d_{fiber}$ from

DDE signal modulation, which allowed distinguishing neuronal and astrocytic compartments (Shemesh et al., 2017): estimated cell-fiber diameter was smaller for NAA ($d_{fiber}$=0.04 µm) than for Ins ($d_{fiber}$=3.1 µm), which was well in line with d values inferred from high-b measurements in the study by Palombo *et al* (Palombo et al., 2017).

In this study, we revisit the works of Shemesh *et al* with two main methodological differences: i) conventional radio frequency (RF) pulses are used (instead of polychromatic RF pulses targeting a few resonances of interest) to detect more metabolites, and ii) state-of-the-art post-processing pipeline is applied for accurate diffusion-weighted signal quantification, similarly to all our previous works on DW-MRS, including scan-to-scan phase correction, LCModel analysis and macromolecule signal quantification. In the end, signal modulation can be reliably quantified for six metabolites. Analysis of the experimental data with a model of isotropically oriented cylinders of infinite length, similarly as in (Shemesh et al., 2017), strongly suggests that this model does not satisfactorily account for our data. We propose that additional structural features are required to better describe DDE-MRS data.

## 2. Methods

### 2.1. DW-MRS experiments

All experimental protocols were reviewed and approved by the local ethics committee (CETEA N°44) and submitted to the French Ministry of Education and Research (approval: APAFIS#795-2015060914444077 v1). They were performed in a facility authorized by local authorities (authorization #B92-032-02), in strict accordance with recommendations of the European Union (2010-63/EEC). All efforts were made to minimize animal suffering and animal care was supervised by veterinarians and animal technicians. Mice were housed under standard environmental conditions (12-hour light-dark cycle, temperature: 22±1°C and humidity: 50%) with ad libitum access to food and water.

Four C57BL/6J wild type mice were anesthetized with 1.5% isoflurane in air/O2 mixture and scanned on an 11.7 T scanner (Bruker, Ettlingen, Germany) with maximal gradient strength on each axis $G_{max}$=752 mT/m, using a cryoprobe. DDE-MRS was performed in a 63 µL voxel positioned around the hippocampus, using a sequence comprising a double spin-echo module with two diffusion blocks followed by a LASER localization module (TE_SE/TE_LASER=119/25 ms, Δ/δ/TM=30/4.5/29.5 ms) similarly to (Shemesh et al., 2014). A total diffusion-weighting of 20 ms/µm² was applied, i.e. b=10 ms/µm² per diffusion block. Water signal was suppressed using a VAPOR module. MM acquisition was performed using a double inversion recovery (TI1/TI2=2200/730 ms) and a diffusion weighting of b=10 ms/µm² on each block.

The first pair of diffusion gradients was applied along X whilst gradient orientation for the second diffusion block was incremented from φ=0 to φ=360° in the XY plane by 45° steps (φ being defined following the convention proposed by Shemesh *et al* (Shemesh et al., 2016)). For each φ value, a total of 128 repetitions (split in four blocks of 32 repetitions interleaved with other φ values) were acquired.

Signal post-processing was performed as described in (Ligneul et al., 2017), including individual scan phasing and inclusion of an experimental macromolecule (MM) spectrum in the LCModel basis set (sum of two experimentally measured spectra). Signal attenuation reported in this paper corresponds to the ratio S(φ,b=20 ms/µm²)/S(b=0.02 ms/µm²).

2.2. DDE data modeling

Experimental data for each animal and each metabolite was first fitted to the phenomenological equation A+B×cos(2φ) as in (Shemesh et al., 2014) using a Monte Carlo approach. The residual sum of squares corresponding to the best initial fit was used as standard deviation to randomly induce artificial Gaussian noise in our experimental data before repeating the fitting operation. This process was performed 10000 times. The resulting A and B coefficients therefore correspond to the mean values obtained over the generated dataset. The ratio $\frac{B}{A}$ was then used to quantify the amplitude of the signal angular modulation with respect to the global MR signal attenuation.

Furthermore, simulations based on analytical computation of DDE-MRS signals in infinite, isotropically oriented cylinders were performed using the MISST toolbox (Drobnjak et al., 2010; Drobnjak et al., 2011; Ianus et al., 2013) for $D_{free}$ ranging from 0.1 to 0.5 µm²/ms and $d_{fiber}$ from 0.1 up to 5 µm, and compared to experimental data to estimate the parameter values that best explain the data, as performed by Shemesh et al. (Shemesh et al., 2017). Calculation of least square residuals was used as the fitting cost function. Errors on estimated parameters were calculated using a Monte Carlo procedure, as described above.

3. Results

3.1. DDE signal modulation for brain metabolites

**Figure 1A** shows a series of DDE-MRS spectra acquired in one mouse whereas **Figure 1B** shows LCModel spectral decomposition for $\varphi=0$ in this experiment. Signal attenuation could be reliably quantified for NAA, tCr, Ins, tCho and taurine (Tau) with Cramer-Rao lower bounds <2% for all $\varphi$ values. Lactate resonance was also quantified, however CRLB values of ~10% were obtained. **Figure 2** displays the signal angular dependence at TM=29.5 ms for six metabolites as well as for MM. In this figure, the experimental data points are fitted to the function $A+B*\cos(2\varphi)$ appearing in light grey. The very low standard deviations that were achieved for all experiments can be appreciated. These results are coherent with Özarslan's work that confirmed Mitra's findings and proposed a Taylor expansion of the MR signal attenuation up to the fourth-order term, showing that the angular dependence of the diffusion signal in finite cylinders can be characterized by the function $A+B*\cos(2\varphi)$ whose coefficients can be related to L and d (Ozarslan, 2009).

Coefficients A and B are reported in the top part of **Table 1**. It can be seen that the amplitude modulation of the DDE-MRS signal - reflected here by B - is larger for NAA (neuronal metabolite) than for non-specific (tCr and Tau) or glial markers (tCho and Ins), which might at first glance be intuitively interpreted as neurons exhibiting narrower fibers, i.e. larger CSA than glia (Palombo et al., 2017; Shemesh et al., 2017), although this simple picture will be challenged later in the following sections. Interestingly, lactate exhibits stronger signal attenuation, which could be explained by the intrinsically fast diffusion of lactate (which is a small metabolite), but also by the contribution of a significant extracellular lactate pool with a rather faster, "free-like" diffusion (Pfeuffer et al., 2000). In line with this idea, lactate exhibits less pronounced angular amplitude modulation than intracellular metabolites, consistent with the idea that, unlike within cellular fibers, in the extracellular space correlation between subsequent diffusion directions is rapidly lost. Lastly, it can be observed that MM signal is little attenuated and does not display any angular dependence, which is consistent with the fact that MM signal results from a pool of slowing diffusing molecules. The normalized modulation $\frac{B}{A}$ is also reported for each metabolite in **Table 1,** and will be used later for comparison with simulations.

3.2. Estimating fiber diameter: simulations in infinite cylinders using MISST

Heatmaps based on least square residuals for each metabolite as compared to DDE simulations performed with MISST, as a function of $D_{free}$ and $d_{fiber}$, are presented in **Figure 3**. They show similar aspect for all metabolites except lactate. Diameters and $D_{free}$ values corresponding to the best fits for each metabolite are reported in the bottom part of **Table 1**. Values for $d_{fiber}$ are in the 3-5 µm range, while $D_{free}$ is found to be ~0.2 µm²/ms. These values are not in good agreement with previously published values. DW-MRS works performed at high b or short $t_d$ rather reported diameters below 3 µm (Kroenke et al., 2004; Ligneul et al., 2017; Marchadour et al., 2012; Palombo et al., 2017) which better agrees with histological estimates for axons and dendrites as well as for astrocytic processes (Ligneul et al., 2019). While it is more difficult to compare $D_{free}$ to ground truth value, past DW-MRS works consistently estimated $D_{free}$ to be larger than 0.3 µm²/ms (and up to ~0.6 µm²/ms for certain works) (Kroenke et al., 2004; Ligneul et al., 2017; Marchadour et al., 2012; Palombo et al., 2016; Palombo et al., 2017; Ronen et al., 2013).

These unexpected results led us to further investigate the effect of d on the amplitude of the signal angular dependence. We therefore carefully examined MISST simulations, in particular for $D_{free}$ ranging from 0.3 to 0.45 µm²/ms and $d_{fiber}$ from 0.5 to 2 µm and subsequently fitted each of the basis signals to $A + B \times \cos(2\varphi)$, as performed for the metabolites. **Figure 4** displays the A, B and $\frac{B}{A}$ for each combination of $d_{fiber}/D_{free}$ values, revealing that no variation of the ratio values was obtained when varying $d_{fiber}$ whereas $D_{free}$ seemed to have a greater effect on the signal angular modulation, therefore illustrating the lack of sensitivity to fiber diameter for $d_{fiber}$<2 µm.

Quite strikingly, $\frac{B}{A}$ values were much higher in the case of the simulated signals (0.26-0.42) than for the metabolites (0.101-0.212). This is well in agreement with the results displayed in **Figure 5,** showing the superposition of NAA and Ins experimental data to a simulated signal obtained from a model of 1-µm diameter infinite cylinders ($D_{free}$=0.35 µm²/ms). Hence, a striking result is that the signal angular dependence of our experimental data (quantified by the ratio $\frac{B}{A}$) is much less pronounced than expected from analytical simulations using realistic parameter values, as least when considering diffusion in randomly oriented infinite cylinders.

## 4. Discussion

### 4.1. Limited sensitivity to cell fiber diameter

Heatmaps of the cost function presented in **Figure 3** exhibit a global minimum for unrealistic $d_{fiber}$ and $D_{free}$ values for all metabolites. In their former study, Shemesh *et al* were able to infer realistic $d_{fiber}$ and $D_{free}$ values, but we were unable to reproduce those results. It should be noted that, although similar diffusion parameters (Δ/δ/TM/b) were used in both studies, there were some notable methodological differences. Shemesh et al. performed acquisitions at ultrahigh field (21.1 T) combined with the use of judiciously designed radiofrequency selective pulses which eliminated the need for water suppression module and simplified spectra, so that signal was quantified by peak integration. Here we worked at lower field strength, but still benefited from exquisite sensitivity thanks to a cryoprobe. We used broadband radiofrequency pulses exciting the whole ppm range to detect more metabolites, and we minimized potential bias in metabolite signal quantification by performing scan-to-scan phase correction, accounting for the contribution of MM signal and using LCModel. As a result, standard deviations achieved in the present study are very low, which might help better identify discrepancies between data and models.

The results shown in **Figure 4** strongly suggest that DDE-MRS offers limited sensitivity to cell fiber diameter, despite the use of a relatively high b-value of 20 ms/µm². Improved sensitivity to $d_{fiber}$ may be achievable by optimizing the gradient waveform, as proposed by Drobnjak *et al.* who were able to distinguish smaller axon radii below 5 µm when using gradients of increasing frequency (Drobnjak et al., 2010). Shemesh later on investigated the DDE and the double oscillating diffusion encoding (DODE) schemes in fixed spinal cords at 16.4 T. He showed superior sensitivity to axon diameter when performing DODE measurements, which is line with the idea that oscillating gradients can probe smaller spin displacements occurring perpendicularly to the main diffusion axis (Shemesh, 2018).

A study performed by Ianus *et al* on DDE and DODE imaging implied that simple geometric models such as infinite cylinders may be insufficient when trying to accurately quantify microscopic diffusion anisotropy and the need of considering structures along the fiber axis was underlined (Ianus et al., 2018). Experimental parameters such as the b-value and sequence timing parameters, in particular $t_d$ and TM, were also pointed out as potential sources of bias when trying to accurately estimate the cell µA.

The unrealistic parameter values extracted from data fitting, as well as the large discrepancy between experimental and simulated data shown in **Figure 5,** clearly suggest that the infinite cylinders model seems to fail to satisfactorily describe our experimental data when using realistic parameter values: in particular, experimentally measured amplitude modulation (and normalized amplitude) appears too low. This suggests that metabolite DDE signature may contain more complex and intricate information reflecting various microstructural features beyond fiber diameter.

4.2. Which feature of cell microstructure could the DDE signal be sensitive to?

In an attempt to better understand what morphological features may explain the observed DDE signal modulation, we considered 3D cell-substrates with additional degrees of complexity. Our driving hypothesis is that the lower signal amplitude modulation observed in the experimental data is due to loss of correlation between subsequent diffusion directions that may be induced by a) non-negligible fraction of diffusing metabolites restricted in isotropic (i.e. spherical) soma compartment and/or b) branching of cellular fibers, leading to hopping of diffusing metabolites from one fiber branch oriented in one direction to another branch oriented in a different direction.

To investigate the validity of these hypotheses, Monte Carlo simulations of diffusion NMR signal were carried out in realistic brain cell morphologies obtained by using the generative model recently introduced by Palombo et al (Palombo et al., 2019). In brief, using 12 selected features like soma radius, branch radius, branch order, branching angle etc., the model generates realistic 3D computational models of any brain cell in a controlled and flexible fashion. Then, the diffusion of 3000 particles within 50 different instances of such realistic cell-substrates (resulting in a total of $1.5 \times 10^5$ diffusing particles) was simulated using Camino (Cook et al., 2006) with diffusion coefficient $D_{free}$ and time step of 20 µs. From the resulting spin trajectories, the DDE signal was computed by phase accumulation approach, using exactly the same sequence parameters of the experiments (see Methods section).

In all simulations described below, eight fibers of finite length were connected to a central "soma". Fiber diameter was fixed to a realistic value ($d_{fiber}$=1 µm), but anyway this has little impact on DDE modulation with the current acquisition parameters, as already explained earlier. The free diffusion coefficient was also set to a realistic value of $D_{free}$=0.35 µm²/ms (Kroenke et al., 2004; Palombo et al., 2017). Starting from that, we investigated the effect of fiber structure (number of successive branches $N_{branch}$ and length of segments between embranchment $L_{seg}$, e.g. as defined in (Palombo et al., 2016) as well as soma size ($d_{soma}$) on DDE behavior. Instead of a systematic study, here we decided to focus on four cases, already spanning a broad range of conditions. To facilitate comparison with intracellular metabolite DDE data, simulation results were also fitted with to $A + B \times \cos(2\varphi)$ to derive A, B and $\frac{B}{A}$ (reported in the top part of **Table 2**).

The first case aimed at assessing DDE behavior for short fibers ($L_{seg}$=30 µm), without successive embranchment ($N_{branch}$ =1), and with a soma being reduced to a simple point ($d_{soma}$=0) (**Figure 6A**). This case is used as benchmark for the simple case of connected randomly oriented thin fibers. A slight decrease of B and $\frac{B}{A}$ can be observed as compared to unconnected infinite cylinders as simulated with MISST, for the same fiber diameter and $D_{free}$.

The second case aimed at evaluating if introducing some branching had a strong effect. Hence we set $N_{branch}$=4, while keeping $L_{seg}$=30 µm and $d_{soma}$=0 (**Figure 6B**). We found that the effect is quite strong, with a large increase in A (becoming actually slightly larger than for experimental data) while B keeps decreasing, resulting in a quite low $\frac{B}{A}$ of ~0.16, i.e. twice as low as the reference case of unconnected infinite cylinders, and much closer to experimentally measured behavior.

We then wondered if a realistic soma, rather than a branched fiber structure, was also able to better account for DDE behavior. Hence we set $d_{soma}$ =10 µm, while imposing a very simple fiber structure, so we set $N_{branch}$=1 and $L_{seg}$=450 µm, so that fibers can be considered "almost" infinite and simultaneously ensures that the soma occupies ~16% of the total cellular volume, which is a realistic volume fraction for both neurons and astrocytes (Chklovskii et al., 2002; Chvatal et al., 2007; Ligneul et al., 2019; Sherwood et al., 2004) (**Figure 6C**). A very strong drop of A is observed, while B is decreased as the same value as case 2. As a result $\frac{B}{A}$ is very high, almost as high as for unconnected infinite cylinders. Overall this is not consistent with experimental data.

The last case we studied was the most realistic (**Figure 6D**), i.e. with the branched fiber structure of case 2 ($N_{branch}$=4, $L_{seg}$= 30 µm), and the realistic soma of case 3 ($d_{soma}$=10 µm). With such parameters the soma still occupies ~16% of the total cellular volume. The resulting behavior is just slightly different from the case with $d_{soma}$=0 (case 2), and is now even closer to experimental data (differences between simulated data and values averaged over intracellular metabolites for A, B and $\frac{B}{A}$ are less than ~25%).

Conversely, we wondered how DDE signals in these various cases would translate into fiber diameter $d_{fiber}$ and $D_{free}$, if "naively" analyzed using the infinite cylinder model as initially done. These results are reported at the bottom of **Table 2**. It clearly appears that branching results in strong underestimation of $D_{free}$ and moderate overestimation of $d_{fiber}$, while the presence of soma has little effect on $D_{free}$ but results in strong overestimation of $d_{fiber}$. In the end, data simulated with the full model (case 4) and fitted with the infinite cylinder model yields $D_{free}$~0.20 µm²/ms and $d_{fiber}$~3.2 µm, which is in surprisingly good agreement with values obtained for intracellular metabolites (average over intracellular metabolites: $D_{free}$~0.20 µm²/ms and $d_{fiber}$~3.8 µm).

Overall these simulations strongly suggest that a branched fiber structure is paramount to get a more sensible interpretation of our DDE-MRS data. A likely explanation for this could be that, in such structure, and considering that segment length is short enough compared to the diffusion distance traveled during TM, the main diffusion direction may change between the two successive diffusion blocks for a significant fraction of molecules, resulting in correlation loss and ultimately reduced amplitude modulation. Adding a non-zero cell diameter seems to better account for experimental data, but considering the sequence parameters used here, the main effect arises from the fiber structure. Although we did not investigate that in the current work, it is possible that fiber undulations would result in a similar effect to that of fiber branching (Brabec J., 2019; Ozarslan et al., 2018).

<u>4.3. Can we glean additional information by performing measurements at short TM?</u>

Both our simulation and experimental data at long TM implied that sensitivity to the microstructure of cell bodies is relatively limited. This is coherent with Mitra's theory that indicates that DDE experiments are only sensitive to the µA arising from the delineations of spherical compartments when the null/short TM condition is met (Mitra, 1995). This was also experimentally confirmed *in vivo* by Koch and Finsterbusch who achieved DDE experiments on a 3T clinical scanner to assess cell size in the human brain (Koch and Finsterbusch, 2008). In this case, the expected signal angular modulation describes a bell-shaped function and the difference measured between the maximum and minimum signal intensities respectively obtained in the anti-parallel ($\varphi$=180°) and parallel cases ($\varphi$=0/360°) can be analytically related to the apparent compartment size (Finsterbusch, 2011; Mitra, 1995).

We therefore postulated that some sensitivity to soma size could be retrieved when acquiring DDE-MRS signatures at shorter TM, and we performed another series of acquisitions for TM=5.5 ms, keeping other parameters unchanged ($\Delta/\delta$=30/4.5 ms, b=10 µm²/ms on each block). In this case, five metabolites could be reliably quantified (NAA, tCr, Ins, tCho and Lac) with CRLB values of 1% for NAA, tCr and tCho, 3% for Ins and 10% for Lac. **Figure 7** displays the angular modulation obtained for these metabolites at TM=5.5 ms (as compared to TM=29.5 ms for reminder). At shorter TM, sinusoidal signal angular modulation could still be observed. More interestingly, a maximum MR signal intensity was reached for $\varphi$=180° while slight signal attenuation was obtained for the $\varphi$=0/360° data points. This trend was particularly remarkable for tCho but could also be observed for NAA, Ins and Lac.

This may indicate an intermediary state between the long and short TM regimes predicted by Mitra and we thus hypothesized that the ratio $\frac{S_{\varphi=180}}{S_{\varphi=0}}$ could possibly reflect an increasing sensitivity to cell bodies when shortening TM. This ratio is reported for the five metabolites at both TM in **Table 3,** consistently reporting higher values in the case of TM=5.5 ms as compared to TM=29.5 ms. These results shall however be interpreted cautiously, as we were able to report a statistically significant difference in the case of tCho only, according to a Kruskal-Wallis test. We further investigated this intuition by performing simulations at TM=5.5 ms in the four models presented in **Figure 7.** This enabled us to compare the evolution of the $\frac{S_{\varphi=180}}{S_{\varphi=0}}$ ratio when shortening TM for different features of the cell morphology. Results are presented in **Table 3** and revealed that the highest increase in the $\frac{S_{\varphi=180}}{S_{\varphi=0}}$ ratio was obtained when incorporating a spherical compartment in our model. In contrast, branching and small fiber lengths seemed to have a negligible effect on this ratio. These findings therefore support the hypothesis that sensitivity to cell body diameter is maximized and might be experimentally accessible at shorter TM.

## 5. Conclusion

DDE measurements of brain metabolites offer an unprecedented and specific access to the CSA of neural and glial cells while requiring relatively low diffusion-weighting. The measured signal carries complex information encompassing several aspects of the cell microstructure such as fiber lengths, branching and soma size, at least in the rodent brain, which mainly consists in grey matter. Mathematical models based on simple geometries for which an analytical solution can be derived, such as the infinite cylinder model, appear insufficient to fully describe experimental data. Our work strongly suggests that an accurate and reliable interpretation of DDE-MRS signature may instead necessitate simulations in complex, realistic 3D cell-substrates. Consequently, a heavy fitting pipeline may be needed to analyze experimental data, for instance involving a dictionary approach based on Monte Carlo simulations (Ligneul et al., 2019; Palombo et al., 2016; Rensonnet et al., 2019). The sensitivity to a specific feature of the cell microstructure, i.e. fiber processes or cell bodies, can be modulated by tuning the DDE sequence timing parameters such as TM. The relationship between sensitivity to a desired microstructural parameter, for instance the fiber diameter, and experimental parameters such as b-value, gradient waveform and diffusion time should be further explored.

## Acknowledgements

This project has received funding from the European Research Council (ERC) under the European Union's FP7 and Horizon 2020 research and innovation programmes (grant agreements No 336331 and 818266, awarded to JV). MP acknowledges support from EPSRC, United Kingdom grant EP/N018702/1. The 11.7 T MRI scanner was funded by a grant from "Investissements d'Avenir - ANR-11-INBS-0011 - NeurATRIS: A Translational Research Infrastructure for Biotherapies in Neurosciences".

# Figures

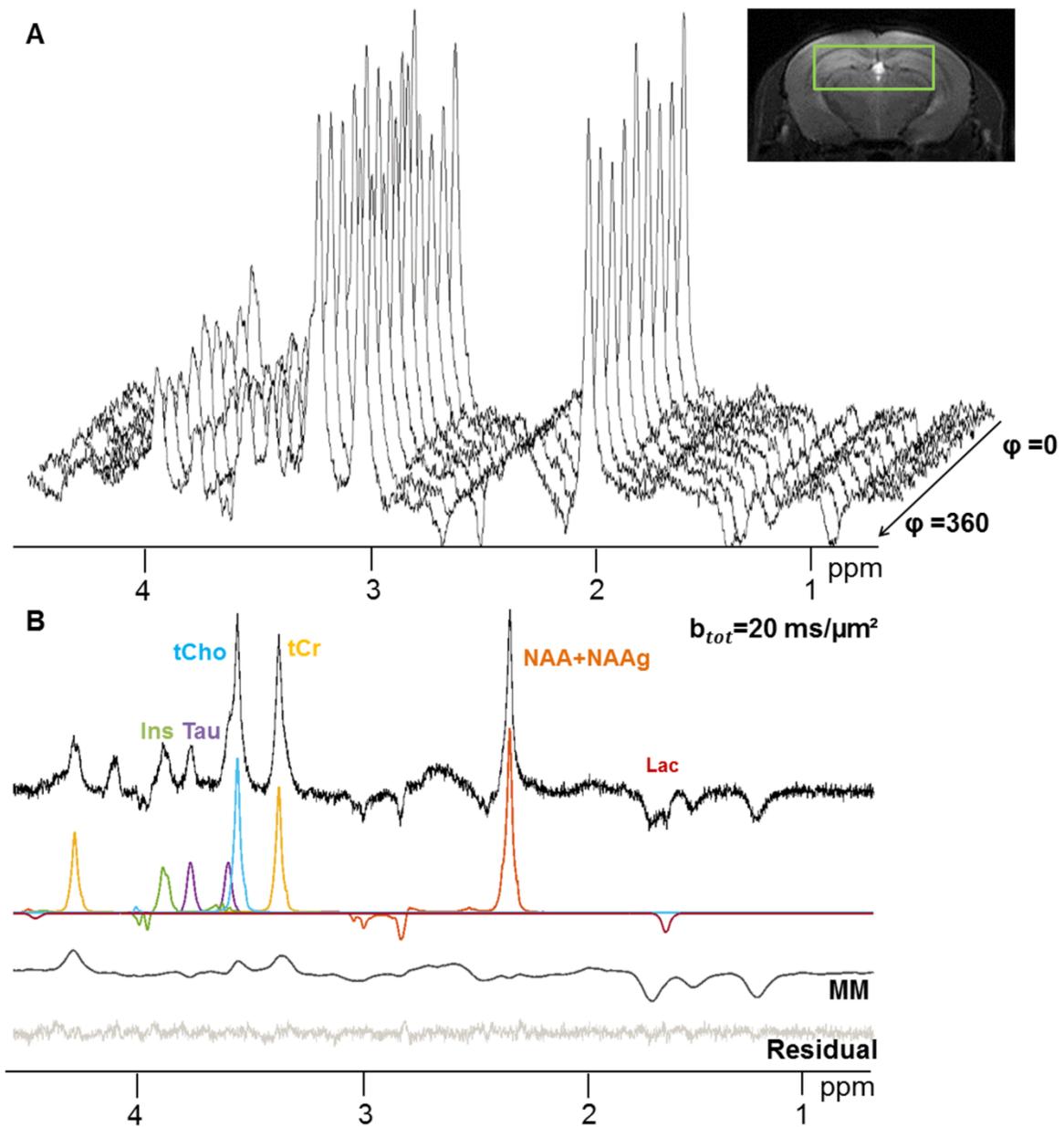

**Figure 1:** Examples of experimental data. **(A)** One dataset acquired using DDE-MRS (Δ/δ/TM=30/4.5/29.5 ms, b=10 ms/μm² per diffusion block), displaying the angular dependency of peak amplitudes. The top-right inset shows the voxel positioned in the mouse hippocampus. **(B)** Spectral decomposition of one diffusion-weighted spectrum using LCModel for 6 metabolites and macromolecules.

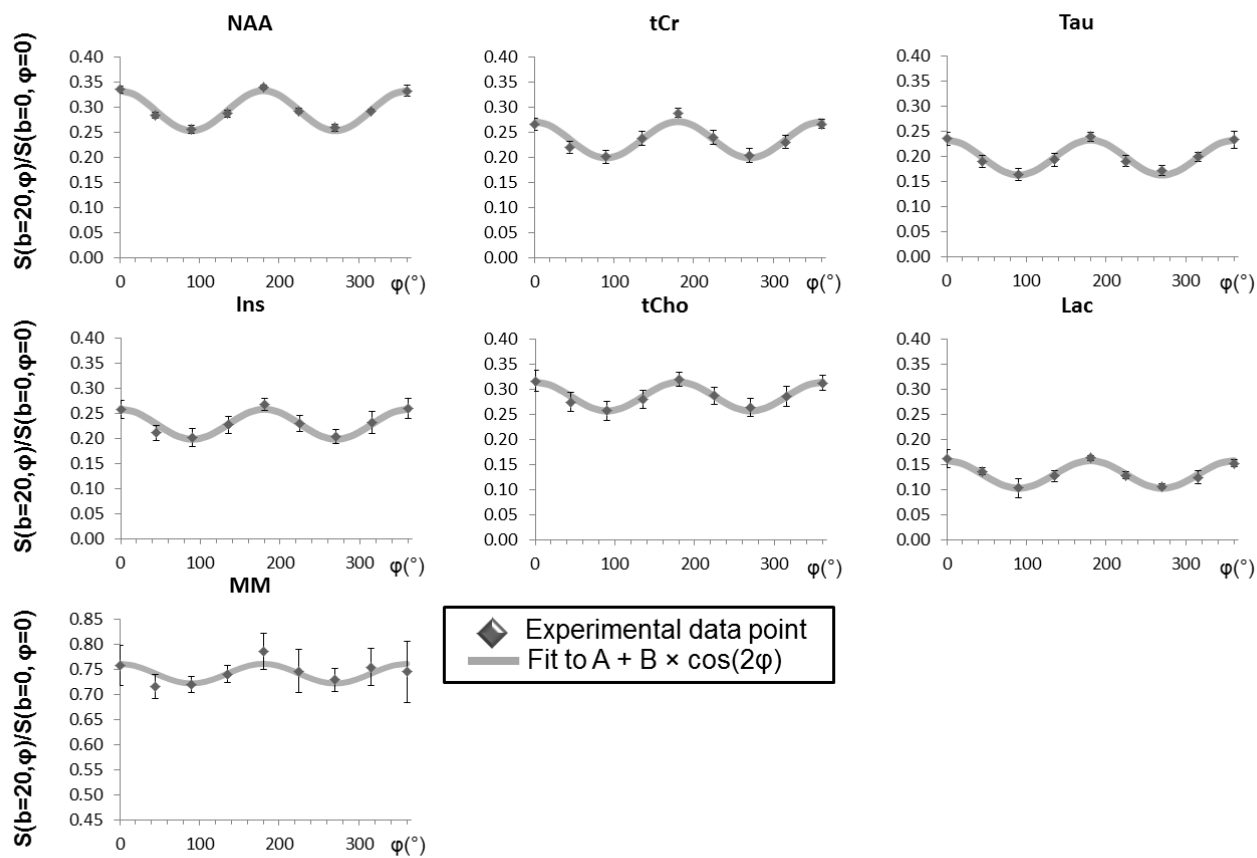

**Figure 2:** Signal modulation in respect to angular values for all 6 metabolites as well as macromolecules. Each experimental point corresponds to the ratio of the signal intensity obtained for b=20 ms/µm² and ϕ= 0-360° to the signal intensity obtained for b=0.02 ms/µm². The gray line corresponds to the function A+B×cos (2ϕ) fitted to the experimental data. Data was acquired in four mice. Error bars stand for the standard deviation.

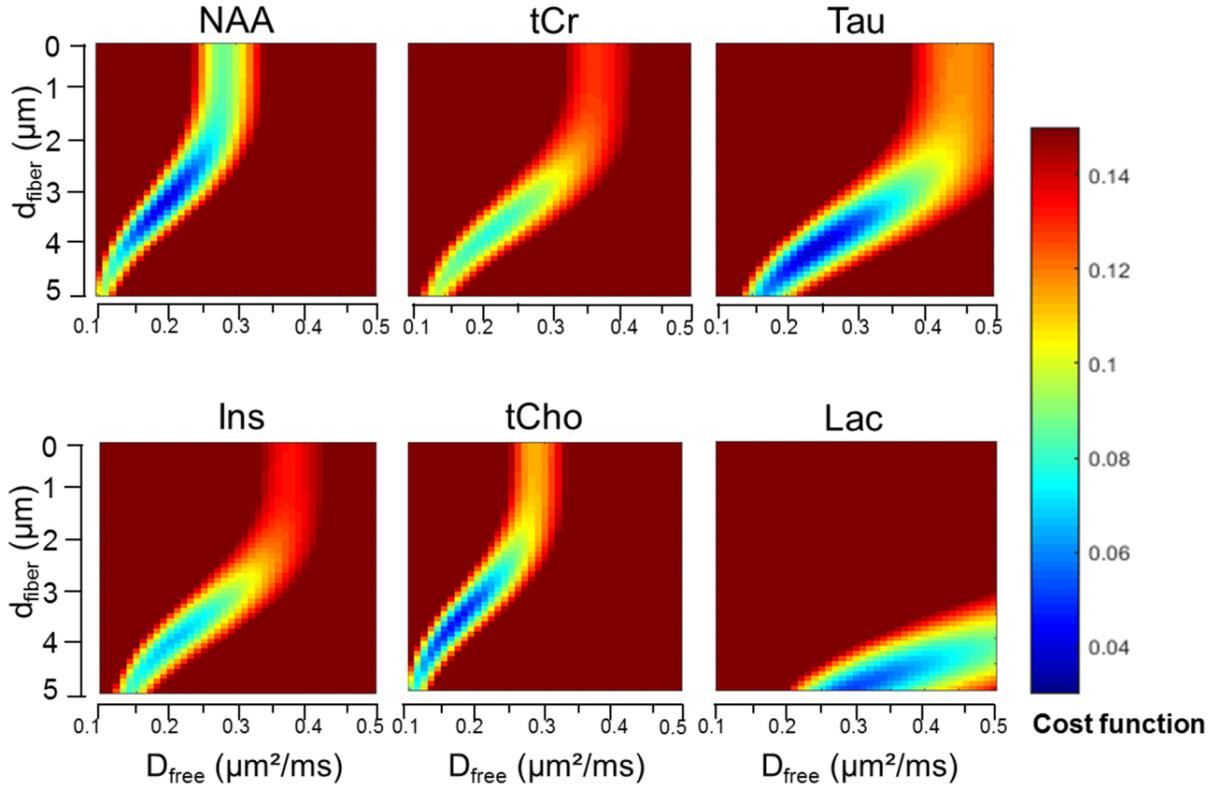

**Figure 3:** Heatmaps of the averaged error estimated for all Dfree (0.1- 0.5 µm²/ms) and $d_{fiber}$ (0.1- 5 µm) values for each metabolite, when using a model of isotropically oriented cylinders simulated by MISST. The colormap represents the cost function $c = \sqrt{\sum_\varphi (x_\varphi - y_\varphi)^2}$, $x_\varphi$ being the simulated diffusion signal attenuation and $y_\varphi$ the experimentally measured signal attenuation for the angle ϕ.

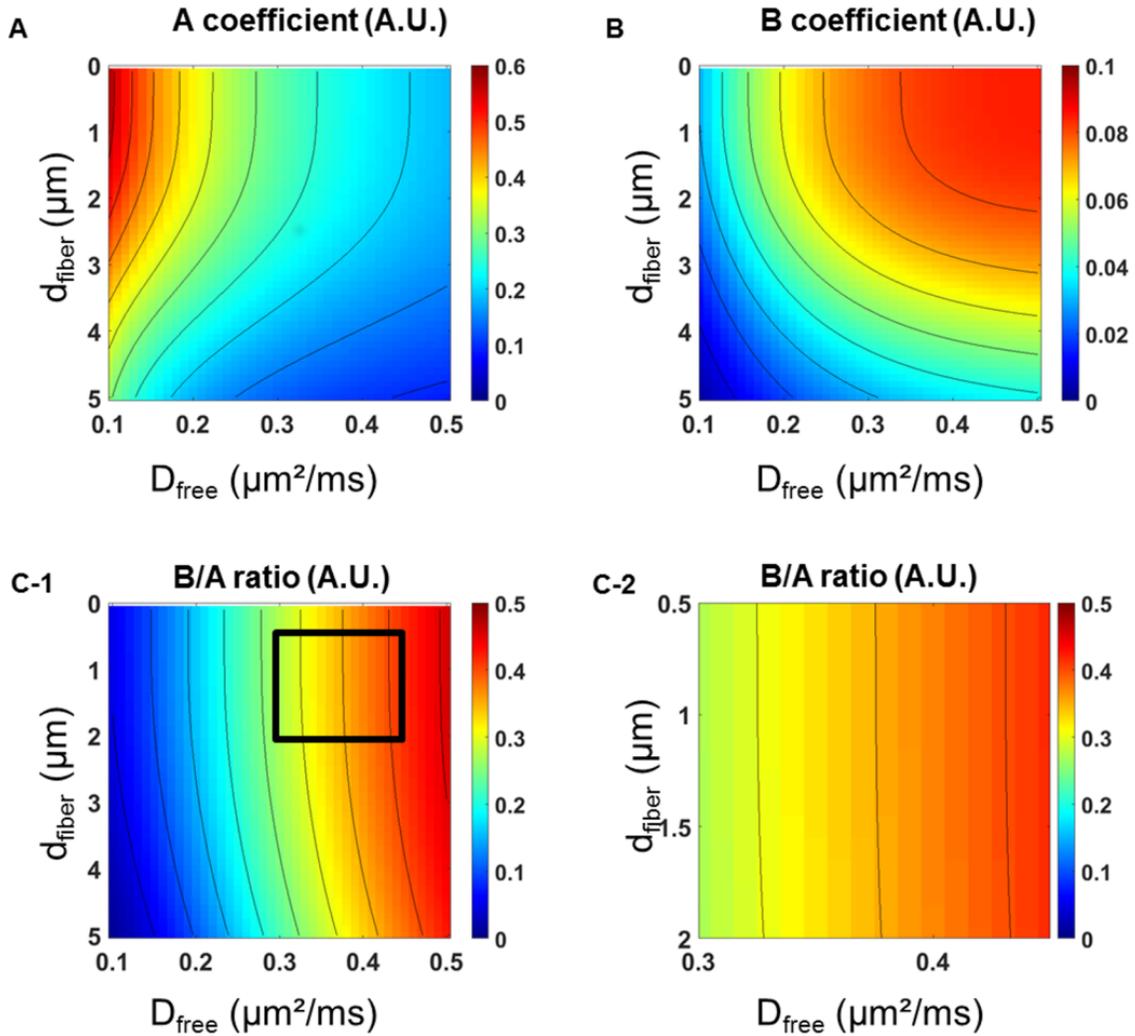

**Figure 4:** A and B coefficients obtained after fitting the simulated dataset in MISST to the $A + B \times \cos(2\varphi)$ function. **(A)** Variations of the A coefficient with respect to $D_{free}$ and $d_{fiber}$. As expected, the overall signal attenuation increases with $D_{free}$ as well as in wider fibers. **(B)** Variations of the B coefficient with respect to $D_{free}$ and $d_{fiber}$. The amplitude of the signal angular modulation is impacted by both $D_{free}$ and $d_{fiber}$, making it challenging to use this parameter to accurately estimate $d_{fiber}$. The sensitivity to $d_{fiber}$ values below 3 µm appears limited, even for a fixed $D_{free}$. **(C-1)** $\frac{B}{A}$ ratios for all $D_{free}$ and $d_{fiber}$ values. **(C-2)** Zoom-in on a realistic range of $D_{free}$ (0.3-0.45 µm²/ms) and $d_{fiber}$ (0.5-2 µm) values, as delineated by the square shown in C-1. The $\frac{B}{A}$ ratios do not vary according to $d_{fiber}$, therefore confirming the poor sensitivity of DDE-MRS experiments to this structural parameter.

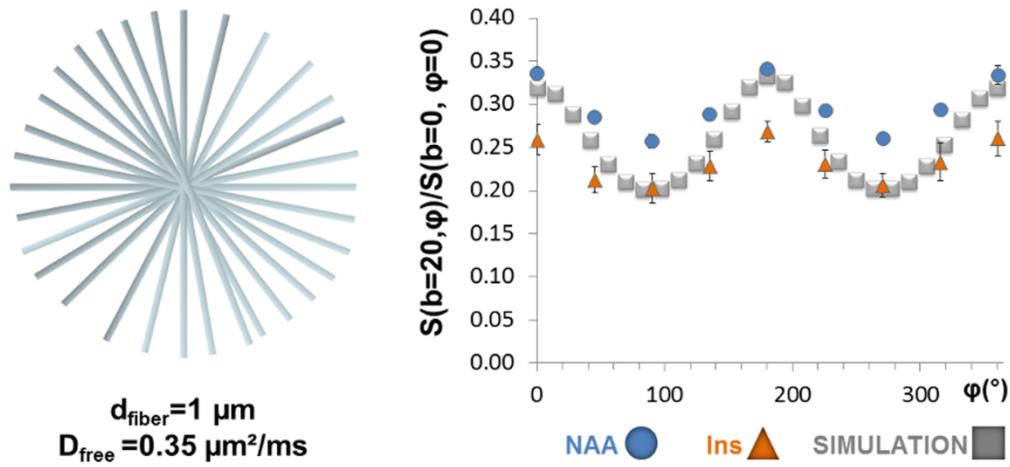

**Figure 5:** Representation of the isotropically oriented infinite cylinders in MISST together with a superposition of a simulated signal obtained for $d_{fiber}=1$ µm and $D_{free}=0.35$ µm²/ms (grey square) to NAA (blue circle) and Ins (orange triangle) experimental data. The amplitude of signal modulation obtained in the simulated case is markedly higher than in the case of our experimental data.

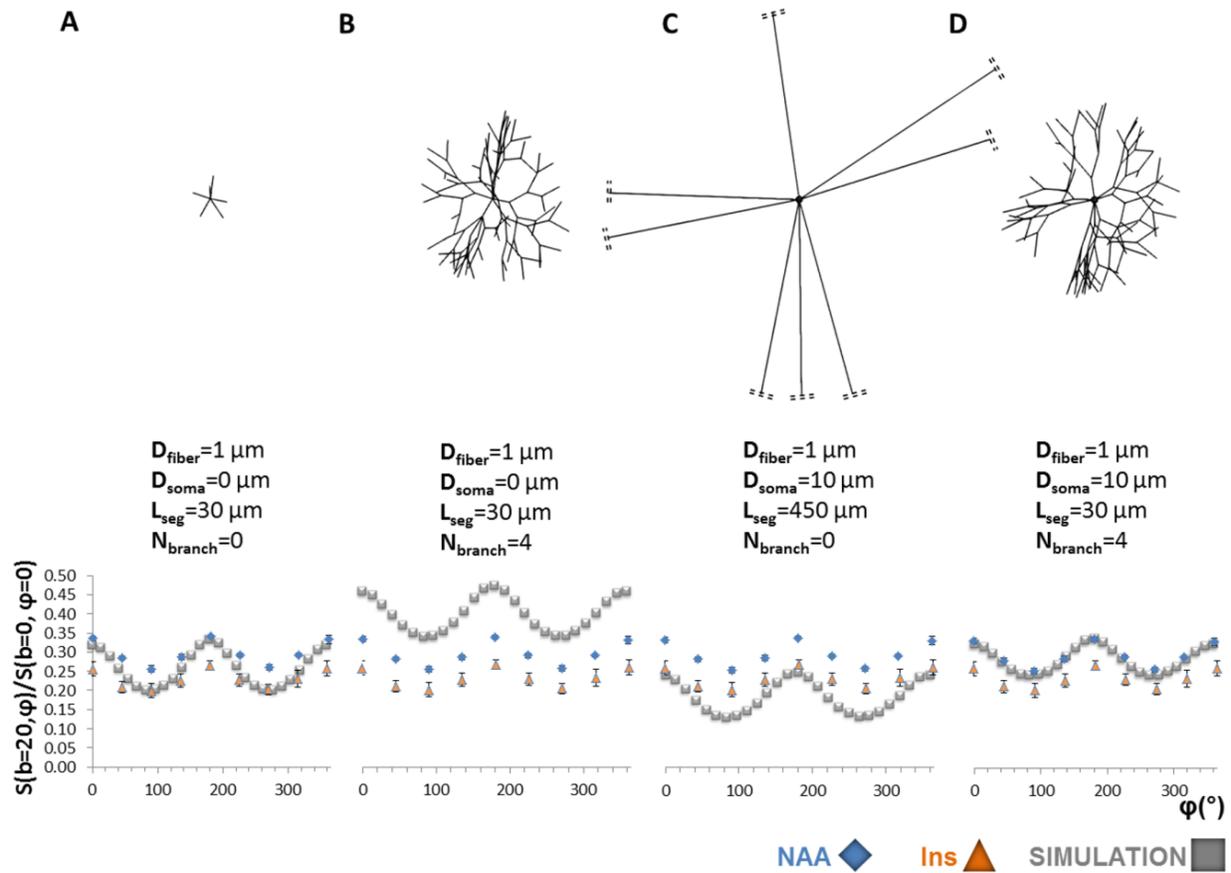

**Figure 6:** Four simulated datasets superposed to NAA and Ins experimental data. **(A)** DDE signal and $\frac{B}{A}$ ratio for a model of short connected fibers. **(B)** DDE signal and $\frac{B}{A}$ ratio for a model of connected fibers exhibiting short segments and branching ($N_{branch}=4$). **(C)** DDE signal and $\frac{B}{A}$ ratio for a model of a 10-µm diameter soma connected to long fibers. The soma volume fraction is ~16%. Note that fibers are cut on the figure (double dash lines). **(D)** DDE signal and $\frac{B}{A}$ ratio for a 10-µm diameter soma connected to fibers with short segments and branching ($N_{branch}=4$). The succession of short segments across successive embranchments appears to best explain our experimental data when comparing $\frac{B}{A}$ ratios. Adding a cell body has little impact on B/A, although it shifts overall signal attenuation (~A) towards lower value (as increased $D_{free}$ would do).

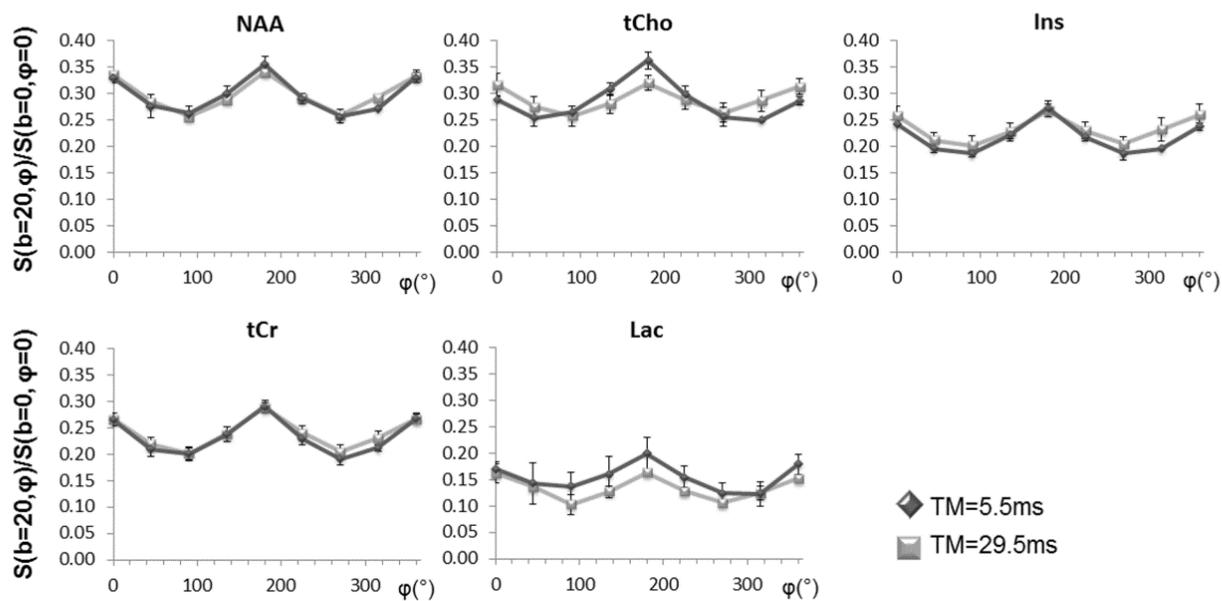

**Figure 7:** Comparison of experimental signal attenuation (mean±s.d. over 4 animals in each group) measured for TM=5.5 and 29.5 ms for 5 metabolites (taurine could not be measured at TM=5.5 ms). For most metabolites it is clear that a peak appears at ϕ=180° when shortening TM.

# Tables

|  | METABOLITES | | | | | |
|---|---|---|---|---|---|---|
|  | NAA (neuronal marker) | tCr | Tau | Ins (glial marker) | tCho (glial marker) | Lac |
| **FITTING TO A + B × cos(2ϕ)** | | | | | | |
| **Offset (A)** | 0.293±0.004 | 0.235±0.012 | 0.197±0.011 | 0.228±0.016 | 0.286±0.018 | 0.130±0.004 |
| **Amplitude Modulation (B)** | 0.040±0.004 | 0.036±0.004 | 0.035±0.005 | 0.030±0.003 | 0.029±0.003 | 0.028±0.006 |
| **B/A ratio** | **0.137±0.014** | **0.155±0.020** | **0.177±0.024** | **0.132±0.015** | **0.101±0.014** | **0.212±0.051** |
| **FITTING TO THE INFINITE CYLINDERS MODEL** | | | | | | |
| **Dfree (µm²/ms)** | 0.20±0.02 | 0.22±0.04 | 0.25±0.03 | 0.21±0.03 | 0.17±0.02 | 0.34±0.06 |
| **d (µm)** | 3.22±0.29 | 3.89±0.43 | 4.20±0.26 | 4.14±0.0.15 | 3.76±0.30 | 4.80±0.27 |

**Table 1:** *Top part:* A and B coefficients obtained for each metabolite when fitting the function A + B × cos (2φ). The B coefficient is higher for neuronal than glial marker. Error corresponds to the standard deviation calculated over four animals. *Bottom part:* $D_{free}$ and $d_{fiber}$ values corresponding to the best fit for each metabolite to a basis of simulated signals using the infinite cylinder model. Error on $D_{free}$ and d was estimated using Monte Carlo simulations. $D_{free}$ appears underestimated whereas $d_{fiber}$ is overestimated.

|  | METABOLITES | GEOMETRIES | | | | |
|---|---|---|---|---|---|---|
|  | Average over NAA, tCr, Tau, Ins and tCho | Infinite cylinders | Short fibers, no soma | Branched fibers, no soma | Long fibers + Soma | Branched fibers + soma |
| **FITTING TO A + B × cos(2ϕ)** | | | | | | |
| **Offset (A)** | 0.248±0.041 | 0.247±0.008 | 0.257±0.006 | 0.353±0.005 | 0.181±0.006 | 0.290±0.006 |
| **Amplitude Modulation (B)** | 0.033±0.005 | 0.081±0.011 | 0.062±0.009 | 0.055±0.007 | 0.055±0.009 | 0.046±0.008 |
| **B/A ratio** | **0.152±0.039** | **0.326±0.045** | **0.242±0.033** | **0.156±0.020** | **0.306±0.050** | **0.158±0.027** |
| **FITTING TO THE INFINITE CYLINDERS MODEL** | | | | | | |
| **Dfree (µm²/ms)** | 0.21±0.03 | 0.35±0.00 | 0.27±0.02 | 0.20±0.01 | 0.33±0.03 | 0.20±0.02 |
| **d (µm)** | 4.00±0.52 | 1.00±0.00 | 2.65±0.39 | 1.89±0.54 | 3.69±0.26 | 3.19±0.29 |

**Table 2:** *Top part:* A, B and $\frac{B}{A}$ values obtained for each model (infinite cylinder model and 3D-cell simulations) when fitting the function A + B × cos (2φ). These values can be compared to the grey column on the left-hand side that indicates the average A, B and $\frac{B}{A}$ calculated over five metabolites (mean ± s.d.). Error on parameters extracted from geometric models ($D_{free}$, d) was estimated using Monte Carlo simulations. Models exhibiting embranchments exhibit lower $\frac{B}{A}$ values that better match our experimental data. *Bottom part:* Fitting to the infinite cylinders model for each 3D-cell simulation. This model fails to yield the expected ground truth values for $D_{free}$ and $d_{fiber}$ (i.e. 0.35 µm²/ms and 1 µm) for all simulations. Most strikingly, the obtained values are close to that of the average $D_{free}$ and $d_{fiber}$ obtained for our experimental data (mean ± s.d. over five metabolites), further illustrating the fact that the infinite cylinders model does not satisfactorily describe the experimental data.

| $\dfrac{S(\varphi = 180°)}{S(\varphi = 0°)}$ | METABOLITES | | | | | GEOMETRIES | | | |
|---|---|---|---|---|---|---|---|---|---|
| | NAA (neuronal marker) | tCr | Ins (glial marker) | tCho (glial marker) ** | Lac | Short Fibers | Branching | Soma | All features |
| TM=5.5 ms | 1.08±0.04 | 1.09±0.06 | 1.14±0.07 | 1.26±0.05 | 1.14±0.12 | 1.07 | 1.04 | 1.16 | 1.12 |
| TM=29.5 ms | 1.02±0.01 | 1.08±0.01 | 1.03±0.02 | 1.02±0.01 | 1.04±0.07 | 1.05 | 1.03 | 1.03 | 1.04 |

**Table 3:** *Left-hand side:* $\dfrac{S_{\varphi=180}}{S_{\varphi=0}}$ ratio obtained for five metabolites at TM=5.5/ 29.5 ms. This ratio tends to increase for all metabolites except tCr at shorter TM, however this increase is statistically significant in the case of tCho only. Error corresponds to the standard deviation calculated over four animals. ** $p<0.05$ according to Kruskal-Wallis test. *Right-hand side:* $\dfrac{S_{\varphi=180}}{S_{\varphi=0}}$ calculated on simulated signals for the four models presented in *Figure 7*. This ratio greatly increases when including a soma in our simulation model, suggesting an increased sensitivity to cell bodies when shortening TM.